\documentclass[%
superscriptaddress,
 amsmath,amssymb,
 aps,
 prx, twocolumn,
longbibliography
]{revtex4-1}
\usepackage{tabularx}
\usepackage{textcomp}
\usepackage{graphicx}%
\usepackage[dvipsnames]{xcolor}

\usepackage{dcolumn}
\usepackage{bm}
\usepackage{color} 
\usepackage{blkarray}
\usepackage{siunitx}
\usepackage{color}
\usepackage{amsmath}
\usepackage{comment}
\usepackage{url} 

\usepackage[colorlinks, citecolor=blue, urlcolor=blue, bookmarks=false, linkcolor=blue, hypertexnames=true]{hyperref} 
\usepackage{soul}

\newsavebox{\overlongequation}

\begin{document}

\title{Vibrational properties of CuInP$_2$S$_6$ across the ferroelectric 
 transition} 


\author{Sabine N. Neal}
\affiliation{Department of Chemistry, University of Tennessee, Knoxville, Tennessee 37996, USA}

\author{Sobhit Singh}
\affiliation{Department of Physics and Astronomy, Rutgers University, Piscataway, New Jersey 08854, USA}

\author{Xiaochen Fang}
\affiliation{Department of Physics and Astronomy, Rutgers University, Piscataway, New Jersey 08854, USA}
\affiliation{Rutgers Center for Emergent Materials, Rutgers University, Piscataway, New Jersey 08854, USA}

\author{Choongjae Won}
\affiliation{Laboratory for Pohang Emergent Materials and Max Plank POSTECH Center for Complex Phase Materials, Pohang University of Science and Technology, Pohang 790-784, Korea}

\author{Fei-ting Huang}
\affiliation{Department of Physics and Astronomy, Rutgers University, Piscataway, New Jersey 08854, USA}
\affiliation{Rutgers Center for Emergent Materials, Rutgers University, Piscataway, New Jersey 08854, USA}

\author{Sang-Wook Cheong}
\affiliation{Department of Physics and Astronomy, Rutgers University, Piscataway, New Jersey 08854, USA}
\affiliation{Rutgers Center for Emergent Materials, Rutgers University, Piscataway, New Jersey 08854, USA}
\affiliation{Laboratory for Pohang Emergent Materials and Max Plank POSTECH Center for Complex Phase Materials, Pohang University of Science and Technology, Pohang 790-784, Korea}

\author{Karin M. Rabe}
\affiliation{Department of Physics and Astronomy, Rutgers University, Piscataway, New Jersey 08854, USA}

\author{David Vanderbilt}
\affiliation{Department of Physics and Astronomy, Rutgers University, Piscataway, New Jersey 08854, USA}

\author{Janice L. Musfeldt}
\email{musfeldt@utk.edu}
\affiliation{Department of Chemistry, University of Tennessee, Knoxville, Tennessee 37996, USA}
\affiliation{Department of Physics and Astronomy, University of Tennessee, Knoxville, Tennessee 37996, USA}

\date{\today}

\begin{abstract}

In order to explore the properties of a  two-sublattice ferroelectric, we measured the infrared and Raman scattering response of CuInP$_2$S$_6$ across the ferroelectric and {\textcolor{black}{glassy}}  transitions and compared our findings to a symmetry analysis,  calculations of phase stability, and lattice dynamics. 
In addition to uncovering displacive character and a large hysteresis region surrounding the ferroelectric transition temperature $T_{\rm C}$, we identify the vibrational modes that stabilize the polar phase and confirm the presence of two ferroelectric variants with opposite polarizations. 
 Below $T_{\rm C}$, a poorly understood \textcolor{black}{relaxational or glassy}  transition at {\textcolor{black}{$T_{\rm g}$}}  is characterized by local structure changes in the form of subtle peak shifting {\textcolor{black}{and activation}} of low frequency out-of-plane Cu- and In-containing modes. The latter are due to changes in the Cu/In  coordination environments and associated order-disorder processes. Moreover, \textcolor{black}{$T_{\rm g}$} takes place in two steps with another large hysteresis region and significant underlying scattering. 
%
%
Combined with imaging  of the room temperature phase separation, 
this effort lays the groundwork for studying 
CuInP$_2$S$_6$ under external stimuli and in the ultra-thin limit.

\end{abstract}

\maketitle
        

\section*{Introduction}

The past decade has witnessed exceptional 
progress in revealing the potential and inner workings of complex chalcogenides, especially  those belonging to the metal phosphorous trisulfide family ($M$PS$_3$, with $M$ = Mn, Fe, Ni) \cite{Burch2018, McGuire2020, Wang2020}. 
Exciting properties under external stimuli include  
sliding, metallicity, piezochromism, and superconductivity under pressure \cite{Haines2018, Wang2018b, Harms2020, Coak2021},  reentrant phases in high magnetic fields \cite{Wildes2020, Wang2021, McCreary2020, Alliati2020}, tunable band gaps \cite{Wang2018a}, and strongly 
anisotropic thermal conductivity \cite{Kargar2020}. 
These systems can also be exfoliated into few- and single-layer sheets that host 
novel magnetic excitations and states as well as symmetry breaking ~\cite{Kaul2014, Huang2017, Kim2019, Long2017, Sun2019, Zhang2015, Kuo2016, Lee2016_nano, Zhong2017,Chu2020}.
While simple metal site substitution is well-studied in the $M$PS$_3$  series, bimetallic 
substitution is relatively unexplored even though lower symmetry may promote useful properties such as ferroelectricity \cite{Singh2008,Swamynadhan2021,Kurumaji2015,Xu2017,Osada2019} along with different types of structural phase transitions. 
Dual-sublattice analogs such as bimetallic CuInP$_2$S$_6$ and  AgInP$_2$S$_6$ offer 
flexible platforms for the discovery of tunable states of matter under external stimuli 
and the development of structure-property relations.  CuInP$_2$S$_6$, for instance, hosts a quadruple-well potential with  two distinct  polar phases and  four different polarization states under strain ~\cite{Brehm2020, NeumayerACSami2020}.

CuInP$_2$S$_6$ is a layered van der Waals system with a paraelectric $\leftrightarrow$ ferroelectric transition near $T_C$ = 310\,K~\cite{Grzechnik998, Simon1994, MAISONNEUVE1995, Fagot_Revurat_2003, Liu2016, Belianinov2015, Dziaugys2020}. 
Although the terms ``ferrielectric" and ``ferroelectric" are both  used in the CuInP$_2$S$_6$ literature \cite{Grzechnik998, Dziaugys2010, MaisonneuvePRB1997, Belianinov2015, Bercha2015, Simon1994, Brehm2020, NeumayerPRA2020},
\textcolor{black}{we have adopted the latter usage here. Our choice is motivated by the observation that there is no In off-centering instability in the absence of Cu off-centering. This contrasts with the case of a ferrimagnet, in which magnetic moments would still appear on either magnetic sublattice even if suppressed on the other sublattice. Here, instead, the small In displacements seem more analogous to the anionic displacements that are also induced by the Cu off-centering.}
In any case, the  polarization in CuInP$_2$S$_6$ is stable and switchable,  although switching under high bias is associated with 
Cu$^+$ ion 
mobility 
\cite{Belianinov2015,Zhang2021}. The latter has a two-step path that involves both in-plane and 
out-of-plane hopping of Cu$^+$ ions \cite{Zhang2021}. 
In the high temperature paraelectric phase, the system is in the $C2/c$ space group, whereas the polar phase has $Cc$ symmetry \cite{Dziaugys2010}.
Individual layers of CuInP$_2$S$_6$ consist of  Cu$^{+}$ and In$^{3+}$ ions surrounded by sulfur octahedra with  P--P dimers filling the octahedral voids. 
Along with the primary driver - which is distortion of cations from their centrosymmetric positions - the symmetry reduction  from $C2/c$ $\rightarrow$ $Cc$ is thought to occur with  ordering of the copper occupancies ~\cite{Dziaugys2010}. This provides a natural  explanation for the high bias ionic conductivity.
Variable temperature Raman scattering  suggests an order-disorder component to the ferroelectric transition as well~\cite{Vysochanskii1998}.

Because $T_{\rm C}$ is just above room temperature, there are a number of intriguing properties at 300 K.
For example, CuInP$_2$S$_6$ displays a room temperature electrocaloric effect that may prove  useful for solid-state refrigeration~\cite{Si2019}. The system also hosts sizable intrinsic negative longitudinal piezoelectricity \cite{Brehm2020,Youeaav3780}. Further, 
CuInP$_2$S$_6$ single crystals exhibit polar domain structure that gets smaller and then disappears in flakes thinner than 50 nm \cite{Belianinov2015,Chyasnavichyus2016,Chen2019}.  CuInP$_2$S$_6$ is also being prepared in thin film form. In fact, when sandwiched with germanene as a two-dimensional van der Waals heterostructure, electric field can drive a metal-semiconductor transition via control over the polarization direction \cite{Zhu2019}. 
 CuInP$_2$S$_6$ has been incorporated (along with MoS$_2$) in a negative capacitance field effect transistor as well \cite{Wang2019}.
 Dielectric dispersion studies  reveal a broad relaxation near 150 K, although the exact position, amplitude, and shape depends upon the measurement frequency \cite{Dziaugys2010}.  This relaxation  is attributed to a dipolar glass transition due to freezing of the electric dipole \cite{Dziaugys2010}. 
Finally, we note that this system hosts a first order monoclinic $\leftrightarrow$ triclinic structural transition at 4 GPa \cite{Grzechnik998}.

In this work, we combine infrared absorption and Raman scattering spectroscopies to reveal symmetry-breaking  and local lattice distortions 
across the  ferroelectric and {\textcolor{black}{relaxational or glassy}} phase transitions in CuInP$_2$S$_6$. We compare our findings with complementary lattice dynamics calculations, mode displacement patterns, and an analysis of the energy landscape.  
Interestingly, we identify several different infrared- and Raman-active modes in this 
system that appear below the ferroelectric transition, although of course only the odd-symmetry  features contribute to the development of electric polarization. The mechanism of the $C2/c$ $\rightarrow$ $Cc$ transition is displacive, and comparison with the $M$PS$_3$ family of materials reveals that bimetallic $A$-site substitution is what enables the polar ground state to emerge \cite{Singh2008,Swamynadhan2021}. 
The 150 K order-disorder transition, on the other hand, does not take place with a change in space group. Yet, it is 
 characterized by subtle peak shifting {\textcolor{black}{ and  activation}} of low-frequency out-of-plane Cu- and In-containing modes that arise due to differences in the local bonding environment. 
 Thus in addition to order-disorder and glassy character to the transition near 150 K, there is a weak structural component due to Cu site disorder.
 The strong hysteresis as well as the two-step character of \textcolor{black}{$T_{\rm g}$} are also consequences of  bimetallic substitution and the resulting Cu site disorder.
We further investigate the tendency toward chemical-phase separation using a combination of vibrational spectroscopies as well as  piezoforce and transmission electron microscopies uncovering the signatures of the highest quality CuInP$_2$S$_6$ crystals and  imaging their two polarization domains. Going forward, these studies will  enable rapid verification of crystal quality.


\section*{Methods}

\paragraph*{Crystal growth and characterization:}  CuInP$_2$S$_6$ single crystals were grown by vapor transport methods as follows. Copper powders, indium shots, a phosphorus lump, and sulfur flakes were loaded into an evacuated quartz tube with $\approx$10$^{-5}$ torr pressure and then heated with a temperature gradient between both ends  of the  quartz ampule. The sealed tube was held at  $\approx$1023\,K for 10 days and then cooled to room temperature. The single crystal flakes were mechanically extracted from the entangled bulk. As detailed in the \textcolor{black}{Supporting Information}, there is 
a strong tendency toward chemical phase separation in CuInP$_2$S$_6$-like materials. In fact, until the crystals are rigorously tested, they should be regarded as Cu$_{1-x}$In$_{1+x/3}$P$_2$S$_6$. Chemical vapor transport growth of CuInP$_2$S$_6$   results  in  thin,  orange,  transparent  flakes. We call these single phase platelets Type I crystals. Type II crystals on the other hand  contain CuInP$_2$S$_6$ +  non-ferroelectric In$_{4/3}$P$_2$S$_6$ as a secondary phase.  All of the spectroscopic work reported here was performed with Type I single crystals.

The exceptional phase complexity makes growth of high-quality stoichiometric crystals quite challenging, and we strongly urge that each crystal be tested for phase purity as discussed in the \textcolor{black}{Supporting Information}. There are a number of tests that can be used to confirm sample quality and purity. In this work,
 we performed energy dispersive x-ray analysis (EDX), piezoforce microscopy, and transmission electron microscopy. 
The vertical piezoforce microscopy (PFM) experiments were performed on freshly exfoliated surfaces of CuInP$_2$S$_6$ using the MultiMode$^{TM}$ atomic force microscope by Veeco/Digital Instruments. Thermally cured gold paste was used to mount the sample and act as the bottom electrode. All PFM measurements were conducted using between 3 and 5 V a.c. The voltage was applied to a conducting contact mode AFM tip, and the bottom electrode was grounded. The vertical piezoelectric response signal was extracted using a NanoScope$^{TM}$ controller and a lock-in amplifier. 
Crystal structure, electron diffraction, and domain character were examined by JEOL-2010F field-emission transmission electron microscopy (TEM) in plane-view specimens. In$_{4/3}$P$_2$S$_6$-containing regions were observed by selecting (002) spots of In$_{4/3}$P$_2$S$_6$.

\paragraph*{Vibrational spectroscopies:} Prior to our spectroscopic measurements, a Type I single crystal of CuInP$_2$S$_6$   was exfoliated to reveal a clean, smooth 
$ab$-plane  
sample surface. This crystal  was adhered to a round pinhole aperture. Infrared measurements were performed in transmittance mode using a Bruker IFS 113V infrared spectrometer equipped with a low noise He-cooled bolometer detector  over the frequency range of 20-700 cm$^{-1}$ with 2 cm$^{-1}$ resolution. The measured transmittance was converted to absorption: $\alpha$($\omega$)= -$\frac{1}{\textit{d}}$ln($\mathcal{T}$($\omega$)), where $\mathcal{T}(\omega)$ is measured transmittance, 
and $\textit{d}$ is the crystal thickness. 
\textcolor{black}{This work was carried out on the $ab$-plane, without polarizers. Note that the $c$-axis is not perpendicular to the $ab$-plane, so there is a residual contribution from the $c$-axis in our spectroscopic results.  This is fortunate, because we were unable to polish a crystal to fully expose the $c$-axis.} 
Raman scattering measurements were performed in \textcolor{black}{a back-scattering geometry} on a Horiba LabRAM HR Evolution Raman spectrometer over a 50-750 cm$^{-1}$ frequency range. We used an excitation wavelength of 532 nm at a power of 0.1 mW, an 1800 line/mm grating, a 50 cm$^{-1}$ notch filter, and a liquid N$_2$ cooled CCD detector. {\textcolor{black}{We used the same orientation of the CuInP$_2$S$_6$  crystal on the pin hole as described above, with light incident onto (and scattered from) the $ab$-plane. No polarizers or analyzers were employed because the instrument does not have them. As a result, the spectra average over the Raman tensor components. Table S1 in the Supporting information summarizes the symmetries of the allowed excitations in the high and low temperature phases of CuInP$_2$S$_6$.} In each case, an open flow cryostat provided temperature control.}

\paragraph*{Electronic structure calculations:} All the first-principles density-functional theory (DFT) calculations were performed using the Projector Augmented Wave (PAW) method as implemented in the Vienna Ab initio Simulation Package (VASP)~\cite{Kresse96a, Kresse96b, KressePAW}. The number of valence electrons in the considered PAW pseudopotentials were 11 ($3d^{10}\,4s^{1}$), 3 ($5s^{2}\,5p^{1}$), 5 (3$s^{2}\,3p^{3}$), and 6 ($3s^{2}\,3p^{4}$) for Cu, In, P, and S atoms, respectively. 
The exchange-correlation functional was computed using the generalized-gradient approximation (GGA) as parameterized by Perdew-Burke-Ernzerhof (PBE)~\cite{PBE}. 
The zero-damping D3 method of Grimme (PBE-D3) was employed to describe the weak van der Waals interactions between the CuInP$_2$S$_6$ layers~\cite{Grimme_D3}. This method has been reported to correctly predict a wide range of physical and chemical properties of CuInP$_2$S$_6$~\cite{Tawfik2018, Reimers2018}. 
The energy convergence criterion for  self-consistent DFT calculations was set  at $10^{-7}$ eV and force convergence criterion for relaxation of atomic coordinates was set at $10^{-3}$ eV/\AA.  
The reciprocal space was sampled using a Monkhorst-pack k-mesh~\cite{MP1976} of size 8$\times$8$\times$4 along with a kinetic energy cutoff of 650\,eV for the plane waves. 

The optimized lattice parameters and cell angles of the paraelectric $C2/c$ phase are $a = b = 6.069$\,\AA, $c = 13.159$\,\AA, $\alpha=\beta=94.5^{0}$, and $\gamma=119.9^{0}$. 
The ferroelectric $Cc$ phase was obtained after  applying a polar $\Gamma_2^{-}$ distortion on the high-symmetry $C2/c$ phase. 
A further free relaxation was performed of the local minimum structure ($Cc$) shown in Fig.~\ref{struct_PE_FE}. The resulting cell parameters and cell angles of the $Cc$ phase are 
$a = b = 6.112$\,\AA, $c = 13.360$\,\AA, $\alpha=\beta=94.3^{0}$, and $\gamma=120.0^{0}$, which are in good agreement with the experimental data reported at 296\,K~\cite{SZhou2021}. 
The {\sc Phonopy} package was employed to calculate the zone-center phonon frequencies and phonon eigenvectors of the DFT optimized structures at 0\,K using the finite-displacement approach~\cite{phonopy}. 
The Bilbao Crystallographic Server was utilized to analyze the symmetry of phonon modes~\cite{bilbao2003}. 
The theoretical infrared (IR) spectra were calculated by computing the mode dynamical charge associated with each
phonon eigendisplacement,  
and the theoretical Raman spectra were simulated by appropriately averaging the Raman activity tensor calculated for each Raman-active phonon eigenmode at zone center~\cite{Skelton2017}. 


\section*{Results and Discussion}

\subsection{Analyzing the symmetries and properties of CuInP$_2$S$_6$}

\subsubsection{Energy landscape and ferroelectricity}

\begin{figure}[t!]
\centerline{
\includegraphics[width = 3.4in]{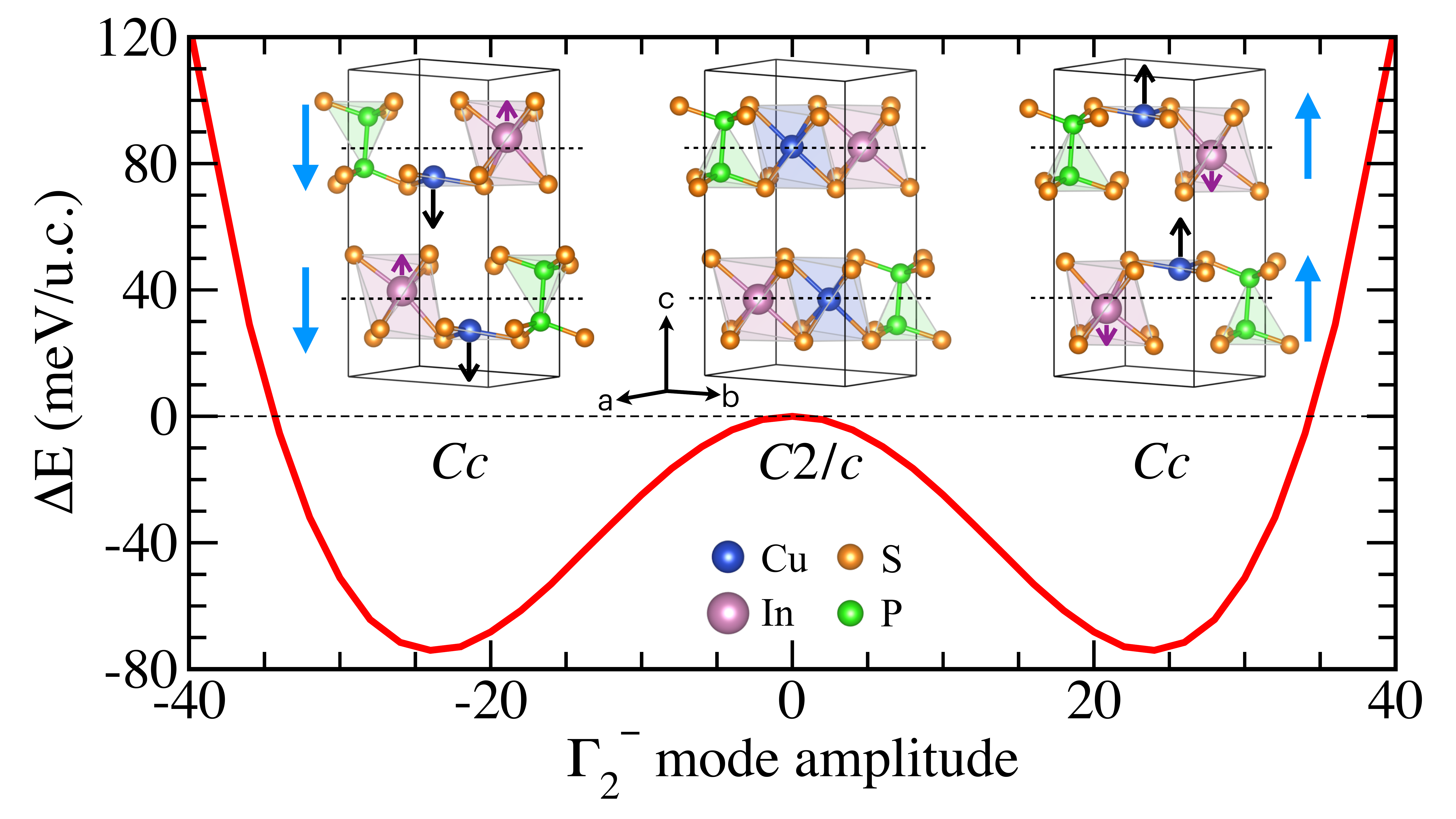}}
\caption{ The double-well potential energy profile computed by freezing the polar $\Gamma_{2}^{-}$ phonon mode in the paraelectric phase as a function of the phonon distortion amplitude. 
Crystal structures of the paraelectric ($C2/c$) and  ferroelectric ($Cc$) phases of CuInP$_2$S$_6$ are shown in the inset.  Black arrows denote the displacement of Cu ions corresponding to the  $\Gamma_{2}^{-}$ mode distortion. Purple arrows denote the relatively small displacement of In ions. The net polarization of individual CuInP$_2$S$_6$ layers is along the direction of the blue arrows. 
}
\label{struct_PE_FE}
\end{figure}

Before we present our  infrared and Raman spectroscopies, let us start by briefly describing the essential details of the crystal structure, ferroelectricity, and vibrational properties of the paraelectric and ferroelectric phases of CuInP$_2$S$_6$. 
The inset of Fig.~\ref{struct_PE_FE} shows the crystal structure of the paraelectric and two equivalent ferroelectric phases of CuInP$_2$S$_6$, which belong to space groups $C2/c$ (no. 15) and $Cc$ (no. 9), respectively. 
The bulk unit cell of both phases contains two weakly interacting single layers of CuInP$_2$S$_6$, $i.e.$, 2 f.u., stacked along the c-axis. 
In the paraelectric phase, the vertically stacked CuInP$_2$S$_6$ layers are related by an in-plane $C_2$ rotation followed by a translation $\tau(c/2)$ operation along the $c$-axis. This structure resembles a lamellar structure composed of a sulfur framework in which the metal cations and P-P dimers fill the octahedral voids within each layer~\cite{Maisonneuve1993, Simon1994, MAISONNEUVE1995, MaisonneuvePRB1997, Fagot_Revurat_2003, Babuka2019}. 
The P-P dimers act as vertical pillars separating the top and bottom sulfur planes in each layer. 
By contrast, the Cu and In metal cations reside exactly at the midplane of the layers, marked by dashed horizontal black lines in the inset of Fig.~\ref{struct_PE_FE}, lying between two vertically stacked sulfur planes and passing through the P-P dimer at its half bond length within each CuInP$_2$S$_6$ layer. 

The paraelectric phase is stable only at high temperatures ($T > 310\,K$)~\cite{Grzechnik998, Dziaugys2010, Belianinov2015, Bercha2015, Simon1994, MAISONNEUVE1995, MaisonneuvePRB1997,Fagot_Revurat_2003}. 
Below 310\,K, this phase transforms into a ferroelectric $Cc$ phase having a net polarization primarily along the out-of-plane direction of each CuInP$_2$S$_6$ layer~\cite{Simon1994, MAISONNEUVE1995, MaisonneuvePRB1997, Fagot_Revurat_2003, Babuka2019}, which occurs due to the polar displacements of the Cu and In sublattices in an antiparallel fashion relative to the midplane of each CuInP$_2$S$_6$ monolayer, as shown in Fig.~\ref{struct_PE_FE}. 

From the group theory perspective, the paraelectric and ferroelectric phases are related by a zone center polar optic phonon mode distortion,  $\Gamma_{2}^{-}$ mode ($B_u$ symmetry). This mode is unstable in the paraelectric phase having frequency $50.3i$\,cm$^{-1}$ and it primarily represents an in-phase vertical displacement of two Cu ions located in the adjacent CuInP$_2$S$_6$ layers, as denoted using two parallel black arrows in the insets of  Fig.~\ref{struct_PE_FE} (left and right panels). 
The calculated potential energy profile obtained by freezing the the $\Gamma_{2}^{-}$ mode as a function of the phonon distortion amplitude is shown in Fig.~\ref{struct_PE_FE}. 
The magnitude of the energy barrier is comparable with the data reported in Refs.~\cite{Brehm2020, NeumayerACSami2020}. Notably, in Refs.~\cite{Brehm2020, NeumayerACSami2020}, the authors reported the presence of two polar phases ($Cc$) with four different (two high and two low) polarization states under local strain conditions. No local strain was applied in our work.  


We find that, in response to the polar displacement of Cu ions ($\sim$\,$1.54$\,\AA), In ions exhibit a relatively small ($\sim$\,$0.20$\,\AA) but nonzero polar displacement in the antiparallel direction to that of the displacement of Cu ions, as shown by purple arrows in the insets of  Fig.~\ref{struct_PE_FE} (left and right panels). 
Such a behaviour of In sublattice has been attributed to the second-order Jahn-Teller effects
~\cite{MaisonneuvePRB1997, Vysochanskii1998, Fagot_Revurat_2003}. 
Although, technically speaking, CuInP$_2$S$_6$ has two polar sublattices, Cu and In, yielding a ferrielectric ordering in each CuInP$_2$S$_6$ layer, only Cu sublattice has a polar instability. In displacement occurs as a response to the polar displacement of Cu ions. \textcolor{black}{In other words, there is no In off-centering instability in the absence of Cu off-centering.} Therefore, it is \textcolor{black}{more natural} to refer CuInP$_2$S$_6$ as a ferroelectric. 



We further compute the magnitude of the net polarization ($\mathbf{P}$) in the $Cc$ phase using the Berry-phase approach~\cite{KSandDV1993, RestaRMP1994}. The theoretically obtained value is $\mathbf{P} = (0.13,\, 0.00,\, 3.43)\,\mu \text{C/cm}^{2}$, which is in excellent agreement with the experimental data reported by Maisonneuve {\it et al.} ($P_z \sim 3.5\,\mu \text{C/cm}^{2}$ at 150\,K)~\cite{MaisonneuvePRB1997}. 
Interestingly, our calculations predict a nonzero in-plane polarization component $P_{x}$, which has not been discussed in the existing literature, to the best of our knowledge.

\begin{figure}[htb!]
\centerline{
\includegraphics[width = 3.4in]{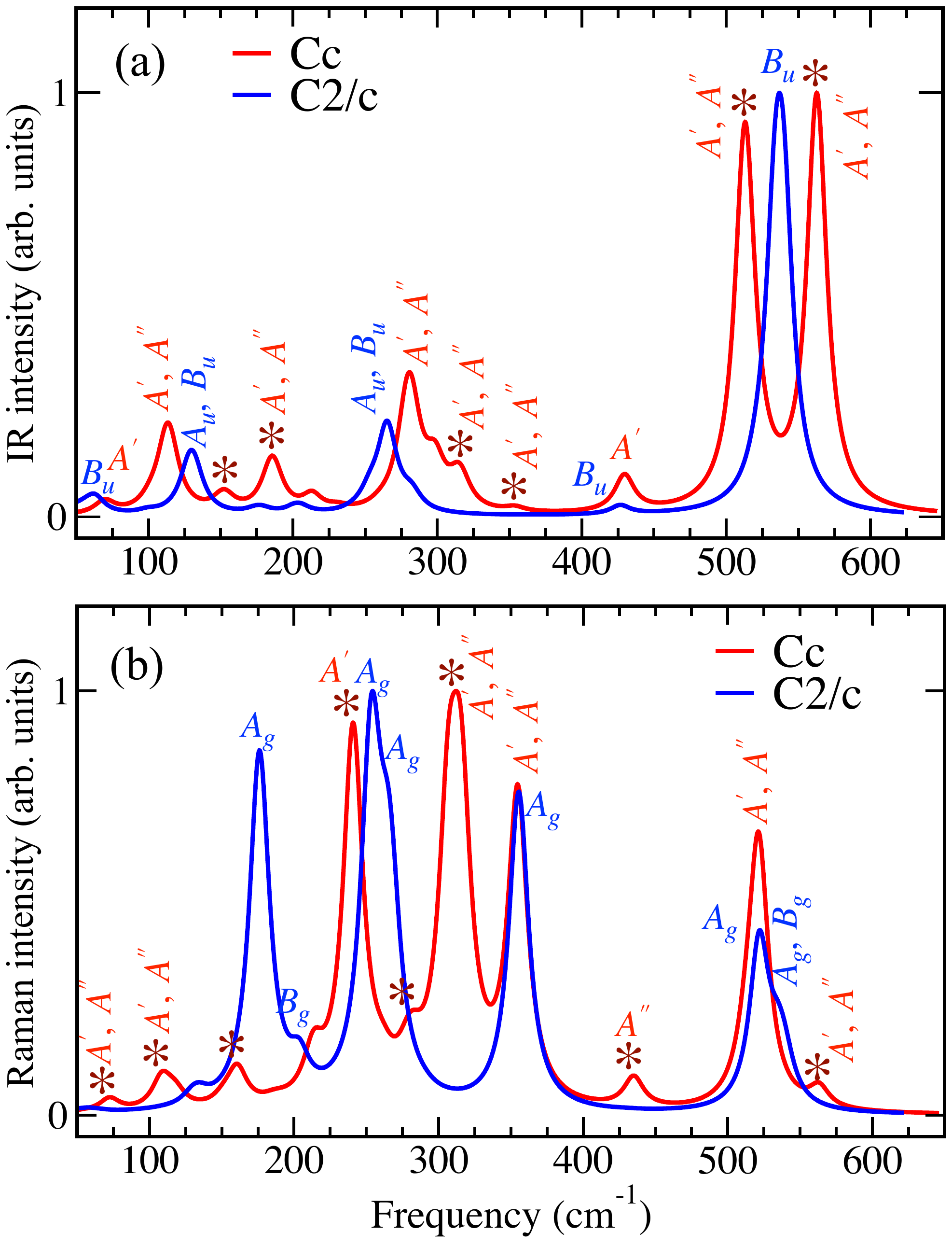}}
\caption{ 
The PBE-D3 calculated (a) infrared and (b) Raman spectra 
for the paraelectric ($C2/c$) and ferroelectric ($Cc$) phases of CuInP$_2$S$_6$ at 0\,K. The symmetries of the major infrared and Raman peaks in the paraelectric and ferroelectric phases are marked in blue and red, respectively. Some peaks are unmarked for clarity. These unmarked peaks contain mixed contribution from both IR active or both Raman active modes in their respective plots, similar to the marked peaks with mixed contributions.   
The \textcolor{BrickRed}{\large{$\ast$}} sign denotes the potential signature modes of the ferroelectric phase, which are absent in the paraelectric phase. The intensities were normalized between zero and one.
\textcolor{black}{A detailed list of each infrared- and Raman-active phonon mode frequency along with its associated atomic-displacement pattern is provided in the Supporting Information~\cite{Supporting}.}  
}
\label{simulated_ir_raman}
\end{figure}

\subsubsection{Selection rules in the paraelectric and ferroelectric phases}

Since the primitive cell contains two formula units, $i.e.,$ 20 atoms/cell, there are sixty allowed phonon modes in CuInP$_2$S$_6$. 
According to the group theory, the allowed acoustic and optical vibrations at zone center can be described using the irreducible representations given below.\\


For the paraelectric phase ($C2/c$): 
\begin{equation}
\begin{aligned}
\Gamma\textsubscript{acoustic} = \emph{A}\textsubscript{u} \oplus 2\emph{B}\textsubscript{u}, \text{~and} \\
\Gamma\textsubscript{optical} = 14\emph{A}\textsubscript{g} \oplus 13\emph{A}\textsubscript{u} \oplus 16\emph{B}\textsubscript{g} \oplus 14\emph{B}\textsubscript{u}.
\end{aligned}
\end{equation}


For the ferroelectric phase ($Cc$):
\begin{equation}
\begin{aligned}
\Gamma\textsubscript{acoustic} = 2\emph{A$'$} \oplus \emph{A$''$}, \text{~and} \\
\Gamma\textsubscript{optical} = 28\emph{A$'$} \oplus 29\emph{A$''$}. 
\end{aligned}
\end{equation}

In the paraelectric phase, only \emph{A}\textsubscript{g} and \emph{B}\textsubscript{g} modes are Raman active while \emph{A}\textsubscript{u} and \emph{B}\textsubscript{u} optical modes are infrared active. 
In the ferroelectric phase, all the \emph{A$'$} and \emph{A$''$} optical modes are both Raman and infrared active. 
Figure~\ref{simulated_ir_raman} shows the simulated infrared and Raman spectra for the paraelectric (red) and ferroelectric (blue) phases. A list of the mode symmetries along with their frequencies is provided in the Supporting Information ~\cite{Supporting}. 
Some distinct features can be noticed in the simulated peak positions of the infrared and Raman spectra which may facilitate the experimental identification of the paraelectric and ferroelectric phases. In Fig.~\ref{simulated_ir_raman}, we marked some Raman and infrared peak positions using \textcolor{BrickRed}{\large{$\ast$}} signs, which denote the potential signature modes of the ferroelectric phase. 
For instance, the $B_{u}$ infrared peak located near $536$\,cm$^{-1}$ converts into two $A'$ and $A''$ peaks located at  $\sim$ 514\,cm$^{-1}$ and $\sim$ 564\,cm$^{-1}$ upon the paraelectric $\rightarrow$ ferroelectric phase transition [see Fig.~\ref{simulated_ir_raman}(a)]. 
Also, a convoluted shoulder consisting of the $A'$ and $A''$ modes develops near $316$\,cm$^{-1}$ frequency in the ferroelectric phase, which is absent in the infrared spectrum of the paraelectric phase. 
\textcolor{black}{Moreover, the unstable polar soft $B_u$ mode ($\Gamma_{2}^{-}$) of the paraelectric phase becomes stable in the ferroelectric phase with a predicted frequency of $\sim$50 cm$^{-1}$ and $A^{\prime}$ symmetry. 
This is also among the signature modes of the ferroelectric phase and should in principle appear in both infrared absorption and Raman scattering. 
This  mode is
near the low frequency limit of our instruments, so we do not unambiguously detect it in our  measurements.  }


\begin{figure*}[tbh]
\centerline{
\includegraphics[width = 6.75
in]{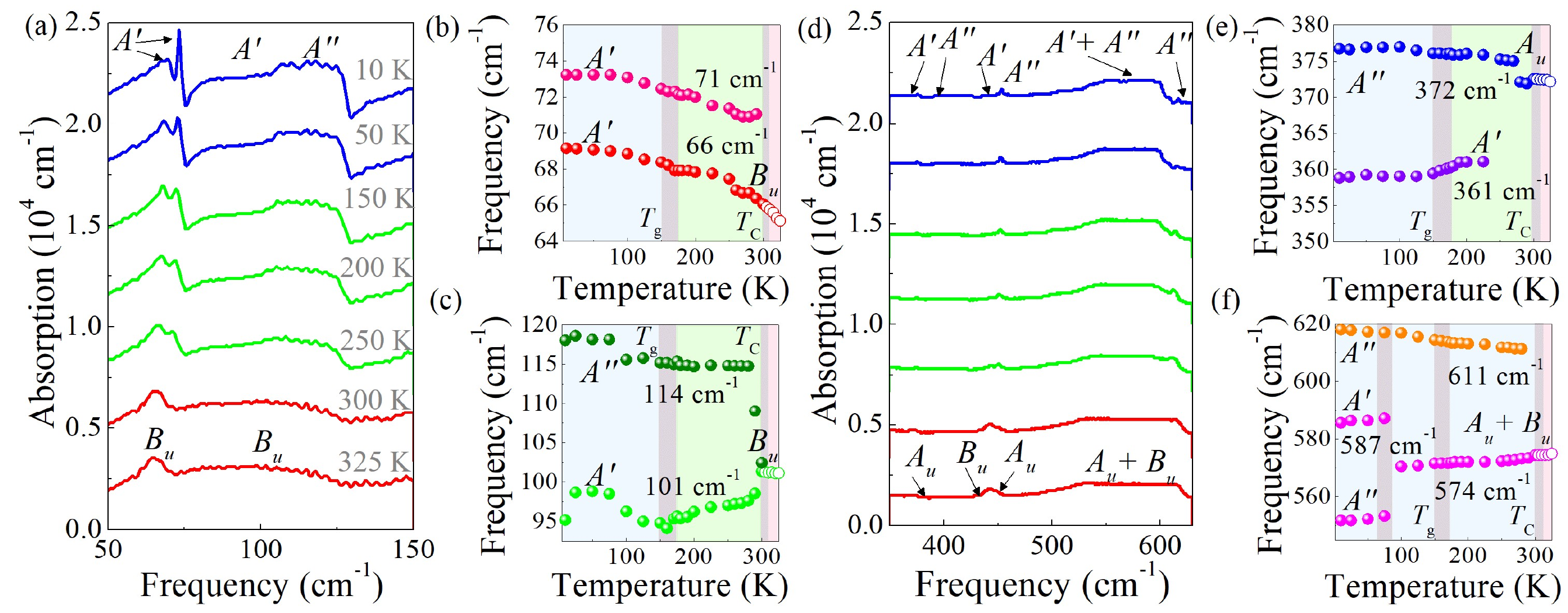}}
\caption{{(a, d) Close-up view of the far infrared response of CuInP$_2$S$_6$ as a function of temperature.  High temperatures are indicated in red, and low temperatures are indicated in blue. The peaks are labeled using $C$2/$c$ and  $Cc$ space group symmetries in the high and low temperature phases, respectively. Panels (b, c) highlight the behavior for the ferroelectric modes in panel (a), whereas  (e, f) display peak position vs. temperature of the ferroelectric modes highlighted in panel (d). The vertical gray bars define the  transition regions, and the open to closed data points denote the $C2/c$ $\rightarrow$ $Cc$ space group change. There is a significant hysteresis depending upon direction of temperature sweep. The measurements shown here are taken with increasing temperature.
}
}\label{Infrared_Tc}
\end{figure*}

On the other hand, the simulated Raman spectrum shows distinct signatures for the paraelectric and ferroelectric phases [see Fig.~\ref{simulated_ir_raman}(b)]. For instance, new Raman peaks are predicted to appear in the ferroelectric phase near frequencies $72$, $110$, $311$, $434$, and $564$\,cm$^{-1}$ which are absent in the Raman spectrum of the paraelectric phase. 
We note that in the ferroelectric phase, except for a few infrared and Raman active peaks, most of the simulated peaks contain contribution from $A^{'}$ and $A^{''}$ modes. This is mainly due to the fact that the  difference in the frequency of most of the $A^{'}$ and $A^{''}$ modes is practically negligible owing to the weak coupling between the adjacent CuInP$_2$S$_6$ layers. 
The simulated Raman spectra of the paraelectric and ferroelectric phases appears to be in reasonable agreement with our experimental data as well as with the data reported by Vysochanskii {\it et al.}~\cite{Vysochanskii1998}. A detailed comparison of the experimental and theoretical data is provided in the 
Supporting Information ~\cite{Supporting}.

\begin{table*}[bth]
\centering
\caption{Infrared-active vibrational modes that are sensitive to the ferroelectric phase transition in CuInP$_2$S$_6$. Frequencies shown for $C2/c$ are at $T$ $\textgreater$ 310 K. For the $Cc$ phase, $T$ $\textless$ 300 K. All values are in cm$^{-1}$. Corresponding DFT calculated frequencies ($\omega$) are given in parentheses.}

\begin{tabular}{||c | c || c |c|| p{5cm} |}
\hline
\multicolumn{2}{|c||}{\textbf{$C2/c$}} & \multicolumn{2}{c|}{ \textbf{$Cc$}} & \\ 
\hline
$\omega$ \textbf{experiment} &  \textbf{symmetry}  &  $\omega$ \textbf{experiment} & \textbf{symmetry}  & \textbf{displacement patterns}\\
\hline\hline

 65 (63) & $B_u$ & - & - & in-plane twist of P-P dimers + out-of-plane vibration of S\\
\hline
- & - & 66 (64) & $A'$ & out-of-plane polar displacement Cu\\
\hline
- & - & 71 (71) & $A'$ & in-plane Cu + In + P + out-of-plane S \\
\hline
 101 (100) & $B_u$ & 102 (102) & $A'$ & out-of-plane rigid shift of P-P dimers (in-phase in adjacent layers) +  out-of-plane vibration of S\\
\hline
- & - & 114 (114) & $A''$ & in-plane Cu + In + out-of-plane S\\
\hline
 - & - & 361 (354) & $A'$ & in-plane Cu + S and out-of-plane P-P stretching \\
\hline
 372 (359) & $A_u$ & 373 (355) & $A''$ & \footnotesize{in-plane Cu + S and out-of-plane stretching of P-P dimers (opposite phase in adjacent layers)} \\
 \hline
 574 (540, 541) & $B_u$, $A_u$ & 553, 587, 611 (558, 563, 563) & $A''$, $A'$, $A''$ & in-plane P-P + in-plane S vibration
vibration\\

\hline
\hline

\end{tabular}
\label{Table 1}
\end{table*}

\subsection{Vibrational properties of CuInP$_2$S$_6$ across the ferroelectric transition}

\subsubsection{Infrared spectroscopy probes inversion symmetry breaking}

Figure \ref{Infrared_Tc} summarizes the infrared response of CuInP$_2$S$_6$ as a function of temperature. In order to analyze the development of polar phonons across $T_{\rm C}$, we focus on the two frequency windows displayed in panels (a, d). 
At 325 K, there are two distinct phonons in the low frequency infrared response. 
These include a sharp $B_u$  mode near 65 cm$^{-1}$ as well as a very broad $B_u$ mode centered around 101 cm$^{-1}$. 
As temperature decreases across $T_C$ = 310 K, there is a noticeable activation of the lower frequency $B_u$ mode and the development of a shoulder on the 101 cm$^{-1}$ feature [Fig. \ref{Infrared_Tc} (b, c)]. The appearance of new modes below 310 K is 
due to a change in the space group from  $C2/c$ in the high temperature paraelectric phase $\rightarrow$ $Cc$ in the low temperature ferroelectric phase. 
While the 65 cm$^{-1}$ feature has $B_u$ symmetry above $T_{\rm C}$,  
the peaks in the low temperature phase are assigned as $A'\oplus A'$. 
This doublet sharpens and continues to blueshift with decreasing temperature. 
At the same time, the once-broad $B_u$ symmetry mode at 101 cm$^{-1}$ develops a shoulder below $T_{\rm C}$. The lower frequency component evolves as an $A'$ mode, and the higher frequency component at 114 cm$^{-1}$ transforms to $A''$ symmetry [Fig. \ref{Infrared_Tc}(c)]. 
The $A'$ branch redshifts below 310 K - a trend that continues to the {\textcolor{black}{relaxational or glassy}} phase 
transition near 150 K  - below which it  
blueshifts again.
By contrast, the $A''$ branch has a sharp blueshift immediately below room temperature and continues to harden systematically to 10 K. 

Figure \ref{Infrared_Tc}(d) summarizes the infrared properties of CuInP$_2$S$_6$ in the 350 to 625 cm$^{-1}$ range. Of the many vibrational modes present in the high temperature phase, only one is sensitive to $T_{\rm C}$: the $A_u$ symmetry mode at 372 cm$^{-1}$ [Fig. \ref{Infrared_Tc}(d, e)]. A slight hardening occurs across $T_{\rm C}$ due to the change in space group, below which the $A_u$ feature converts to an $A''$ symmetry mode. Below $T_{\rm C}$, this mode downshifts slightly before rising near 275 K, continuing to blueshift toward base temperature. Activation of a small $A'$ symmetry mode at 361 cm$^{-1}$ can be seen in the low temperature phase near 225 K, with an overall redshift toward base temperature. Figure \ref{Infrared_Tc}(d, f) also shows a broad phonon centered 574 cm$^{-1}$, assigned as an $A_u \oplus B_u$ symmetry mode. This feature redshifts slightly across $T_{\rm C}$ below which the new symmetries are assigned as $A' \oplus A''$. Systematic redshifting persists through $\approx$100 K, where this broad feature \textcolor{black}{separates} slightly into two individual symmetry components: $A'$ (587 cm$^{-1}$) and $A''$ (553 cm$^{-1}$). 
Both features  redshift slightly to 10 K. The spectra also reveal an $A''$ mode at 611 cm$^{-1}$ that emerges  just below 300 K. The appearance of this feature is in line with our theoretical predictions. This mode sharpens and increases in frequency toward 10 K. A comprehensive table of mode assignments and displacement patterns is available in the Supporting Information. 

\begin{figure*}[tbh]
\centerline{
\includegraphics[width = 6.75
in]{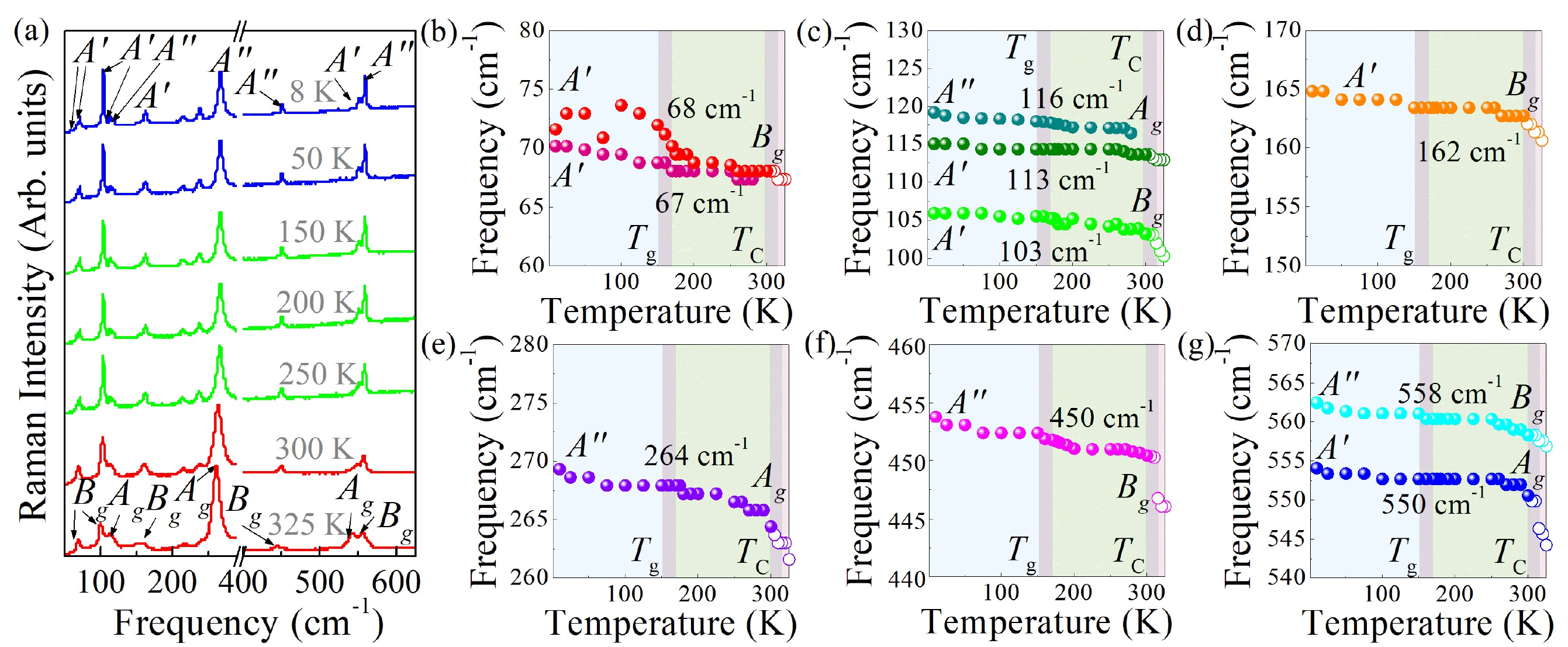}}
\caption{{ (a) Raman scattering  of CuInP$_2$S$_6$ as a function of temperature.   High temperatures are indicated in red, and low temperatures are indicated in blue. The peaks are labeled using $C$2/$c$ and  $Cc$ space group symmetries in the high and low temperature phases, respectively. Panels (b-g) highlight the behavior of several different Raman-active modes, many of which are sensitive to $T_{\rm C}$. The vertical gray bars define $T_{\rm C}$ and \textcolor{black}{$T_{\rm g}$}. The open to closed data points  indicate the  $C2/c$ $\rightarrow$ $Cc$ transition. There is a significant hysteresis effect depending upon direction of temperature sweep. This is seen very clearly in the Raman scattering response when rendered as a contour plot [Fig. S3, Supporting Information]. The measurements shown here correspond to a temperature up-sweep.
}
}\label{Raman_Tc}
\end{figure*}

\begin{table*}[bth]
\centering

\caption{Raman-active vibrational modes that are sensitive to the ferroelectric phase transition in CuInP$_2$S$_6$. Frequencies shown for $C2/c$ are at $T$ $\textgreater$ 310 K. For the $Cc$ phase, $T$ $\textless$ 300 K. All values are in cm$^{-1}$. Corresponding DFT calculated frequencies ($\omega$) are given in parentheses.}

\begin{tabular}{||c | c || c |c|| p{5cm} |}
\hline
\multicolumn{2}{|c||}{\textbf{$C2/c$}} & \multicolumn{2}{c|}{ \textbf{$Cc$}} & \\ 
\hline
$\omega$ (\textbf{experiment}) &  \textbf{symmetry}  &  $\omega$ (\textbf{experiment}) & \textbf{symmetry}  & \textbf{displacement patterns}\\
\hline\hline

67 (66)  & $B_g$ & - & -  & in-plane Cu + P and out-of-plane S vibration \\
\hline
-  & - & 68 (64), 67 (70) & $A'$, $A''$  & out-of-plane Cu +S vibration \\
\hline
 100 (108) & $B_g$  & 103 (102) & $A'$ & out-of-plane In + P + S    \\
\hline
113 (130) & $A_g$  & 114 (114), 116 (117)  & $A'$, $A''$ & in-plane displacement Cu + In + S, out-of-plane S   \\
\hline
 161 (175)  & $A_g$ & - & - & in-plane Cu + P and out-of-plane S  \\
 \hline
 - & - & 162 (160) & $A'$ & out-of-plane P-P + in-plane S  \\
 \hline
 262 (254) & $A_g$ & - & - & in-plane Cu + P + out-of-plane S  \\
 \hline
 - & - & 264 (263) & $A''$ & in-plane S motion\\
 \hline
 446 (429) & $B_g$ & 450 (435) & $A''$ & out-of-plane P + S   \\
 \hline
 544 (534) & $A_g$ & 550 (558, 558)  & $A'$, $A''$ & in-plane P-P stretching + S vibration   \\
 \hline
 557 (539) & $B_g$  & 558 (562, 563)  & $A'$, $A''$  & in-plane P-P + S vibration
\\

\hline
\hline

\end{tabular}
\label{Table 2}
\end{table*}

The development of a polar state requires  inversion symmetry breaking, so we are naturally interested in the behavior of the odd-symmetry vibrational modes across $T_{\rm C}$. Our infrared work demonstrates that CuInP$_2$S$_6$ hosts five polar modes in reasonable overall agreement with the theoretical results in Fig. \ref{simulated_ir_raman}(a). These include features near 65 and 103 cm$^{-1}$ that \textcolor{black}{appear to split below $T_{\rm C}$ due to the activation of new modes}, the  372 cm$^{-1}$ feature which blueshifts across the ferroelectric transition, and two additional features near 361 and 611 cm$^{-1}$ {\textcolor{black}{are activated}}  in the low temperature phase. 
All infrared-active polar modes, along with their symmetries and displacement patterns, are listed in Table \ref{Table 1}.  \textcolor{black}{The predicted $A^{\prime}$ symmetry soft mode near 50 cm$^{-1}$ is not unambiguously observed in our spectra - probably due to the low frequency limit of our instruments.}

\subsubsection{Raman scattering spectroscopy across $T_{\rm C}$}

Figure~\ref{Raman_Tc}(a) summarizes the Raman scattering response of CuInP$_2$S$_6$ along with mode assignments for the various phonons as a function of temperature. Even-symmetry modes also change across $T_{\rm C}$ although they do not break inversion symmetry. 
For instance, the 68 cm$^{-1}$ $B_g$ symmetry mode in the high temperature paraelectric phase \textcolor{black}{transforms}  to give $A' \oplus A''$ symmetry modes in the low temperature ferroelectric phase [Fig. \ref{Raman_Tc}(b)]. 
The upper frequency  branch is relatively consistent in terms of size and position until 150 K, below which the peak position hardens toward 75 K and then softens again. 
The lower frequency branch systematically blueshifts toward base temperature. 
Figure \ref{Raman_Tc}(c) displays the behavior of the $A_g \oplus B_g$ high temperature doublet.  
The $B_g$ component near 103 cm$^{-1}$ hardens  toward $T_{\rm C}$ then remains relatively constant (although the symmetry becomes $A'$). 
Below 300 K, the $A_g$ peak \textcolor{black}{converts to} an $A'$ branch near 113 cm$^{-1}$ and an $A''$ branch near 116 cm$^{-1}$. 
Moving on to the $A_g$ feature at 162 cm$^{-1}$ [Fig. \ref{Raman_Tc}(d)], we note a 4 cm$^{-1}$ blueshift across the ferroelectric  transition  as the mode changes to $A'$ symmetry. The peak position remains relatively constant, shifting only slightly with decreasing temperature. 
Figure \ref{Raman_Tc}(e) displays the frequency vs. temperature trend for the $A_g$ symmetry mode near 264 cm$^{-1}$. It hardens across the $C$2/$c$ $\rightarrow$ $Cc$ transition and is reclassified as $A''$ which has a fairly continuous up-shift to base temperature. 
The blueshift near the ferroelectric transition is even more noticeable in the $B_g$ symmetry mode near 450 cm$^{-1}$ [Fig. \ref{Raman_Tc}(f)]. 
Below 310 K, the feature hardens steadily until 150 K where there is a slight cusp and then an overall blueshift toward base temperature. Finally, we turn our attention to Fig. \ref{Raman_Tc}(g) which tracks the high frequency $A_g \oplus B_g$ doublet. The higher frequency branch has an even clearer blueshift across $T_{\rm C}$ although both features move systematically below 310 K, with a slight frequency upturn  below 25 K. 
A full list of symmetries and mode assignments as well as the relevant mode displacement patterns are available in the Supporting Information.

\begin{figure*}[tbh]
\includegraphics[width = 6.75in]{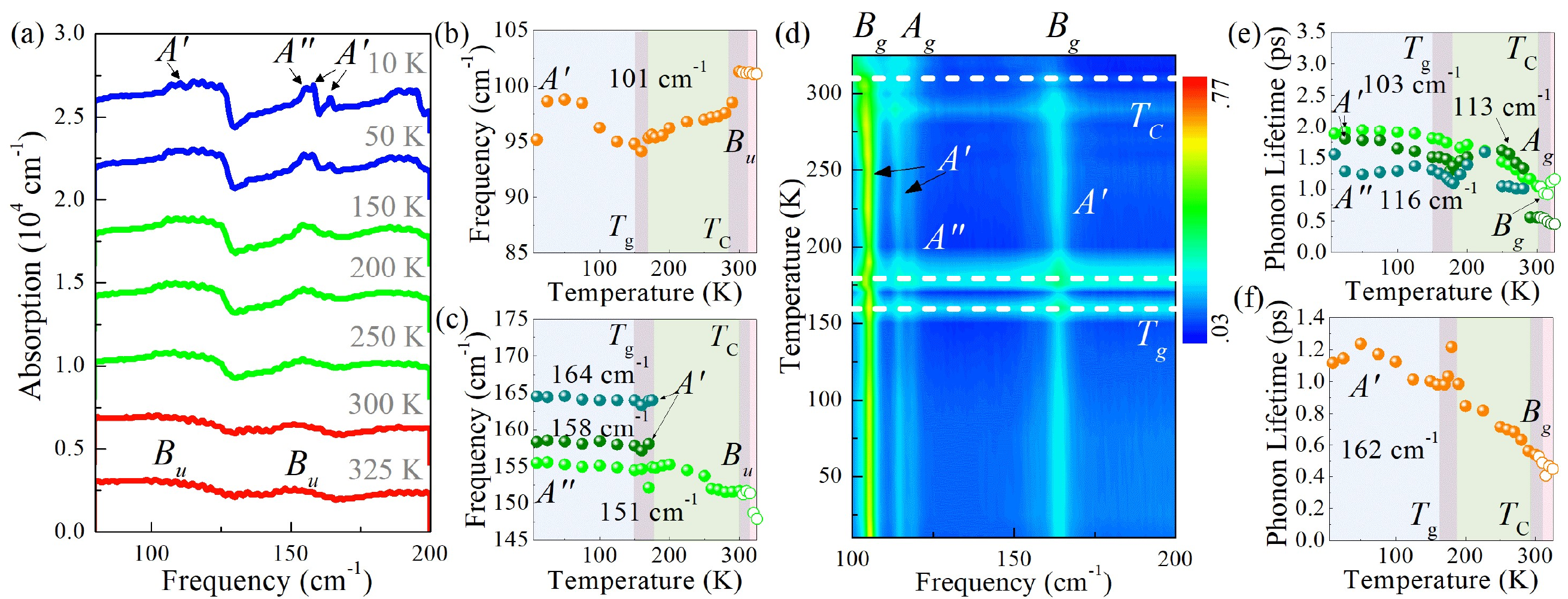}
\caption{{Infrared (a-c) and Raman scattering (d-f) response of CuInP$_2$S$_6$ across the {\textcolor{black}{relaxational or glassy phase transition, $T_{\rm g}$.}}  High temperatures are shown in red, and low temperatures are shown in blue. The peaks are labeled using symmetries of the $C$2/$c$ and  $Cc$ space groups in the high and low temperature phases, respectively.}  Panel (a) summarizes the infrared spectrum as a function of temperature, and panels (b, c) show peak position vs. temperature to highlight the infrared-active modes most influenced by \textcolor{black}{$T_{\rm g}$} . The curves in panel (a) are off-set for clarity. Panel (c) summarizes the variable temperature Raman scattering response as a contour plot, and panels (e, f) deepen our understanding of the local structure transition at \textcolor{black}{$T_{\rm g}$} by examining phonon lifetime trends as discussed in the text. There is a significant hysteresis effect in both $T_{\rm C}$ and \textcolor{black}{$T_{\rm g}$} depending upon direction of temperature sweep [Fig. S3, Supporting Information].  These measurements correspond to a temperature up-sweep. 
}
\label{Ts_infrared_raman}
\end{figure*}

As previously mentioned, several even-symmetry Raman-active  modes change across $T_{\rm C}$ although they do not break inversion symmetry  or contribute to the electric polarization in CuInP$_2$S$_6$. Our Raman scattering work reveals 
{\textcolor{black}{ activation of additional modes related to the}} 67 and 113 cm$^{-1}$ phonons below $T_{\rm C}$. This is in line with theory for a displacive transition, which predicts new peaks in the ferroelectric phase. Other Raman-active phonons including those at 100, 161, 262, 446, and the 544--557 cm$^{-1}$ doublet reveal sharp frequency shifts near  310 K as well. Frequency shifts are also evident in the predicted Raman spectra [Fig. \ref{simulated_ir_raman}(b)]. All Raman-active modes that are sensitive to $T_{\rm C}$, along with their symmetries and displacement patterns, are summarized in Table \ref{Table 2}.
Raman scattering nicely shows the hysteresis across $T_{\rm C}$  as well [Fig. S3, Supporting Information].

\subsubsection{Bimetallic $A$-site substitution introduces ferroelectricity}

Returning to a comparison of $M$PS$_3$ systems ($M$ = Mn, Ni, Fe) and dual sublattice analogs such as CuInP$_2$S$_6$ and AgInP$_2$S$_6$, we immediately see that the most important consequence of the bimetallic nature is the introduction of ferroelectricity. $A$-site size disorder has been predicted produce polar behavior in other materials as well \cite{Singh2008,Swamynadhan2021}. That $T_{\rm C}$ is slightly above room temperature is already  providing the basis for a number applications. Examples include electrocaloric effects for refrigeration,  negative longitudinal piezoelectricity, and a negative capacitance field effect transistor \cite{Si2018, Si2019}.  Unfortunately, Cu site disorder demonstrably broadens the ferroelectric and {\textcolor{black}{relaxational or glassy}} phase transitions.  This is seen very clearly in the Raman scattering response when rendered as a contour plot Fig. 5(d) and [Fig. S3, Supporting Information]. 
It also  hinders the formation of larger polarization in  CuInP$_2$S$_6$.
%
The $M$PS$_3$ materials obviously do not have $A$-site disorder. They also lack ferroelectricity (although MnPS$_3$ may be ferrotorroidic)  \cite{Ressouche2010}. Another useful point of comparison in these structure-property relationships is  CrPS$_4$, a related  chalcogenide with slightly different stoichiometry and $C2$ symmetry \cite{Neal2021}. This system appears to be both polar and chiral.  There is no $A$-site disorder, and the P--P dimer is absent. Clearly, subtle structural changes have important consequences for properties in this family of materials.

\subsection{Vibrational properties of CuInP$_2$S$_6$ across the \textcolor{black}{low temperature relaxational transition} }

In addition to the  well-defined ferroelectric transition, CuInP$_2$S$_6$ displays a weak \textcolor{black}{relaxational or glassy transition ($T_{\rm g}$)} near 150 K.
By comparison, this transition is more subtle and much less studied than the displacive $C$2/$c$ $\rightarrow$ $Cc$ ferroelectric transition. 
Although \textcolor{black}{$T_{\rm g}$} does not alter the $Cc$ space group, there is a local rearrangement of the structure in which the relative inter-atomic coordinates of Cu and In are different. The weak structural component is  
probably due to the existence of multiple competing $Cc$ phases with varying monoclinic cell angle~\cite{SZhou2021}. This gives the transition  local structure aspects as well as order-disorder or glassy character due to freezing of ferroelectric domains \cite{Dziaugys2010}.

Figure~\ref{Ts_infrared_raman} summarizes the infrared absorption and Raman scattering response of CuInP$_2$S$_6$ across the \textcolor{black}{relaxational} transition. We can track how the various modes evolve across $T_{\rm g}$ using frequency vs. temperature and phonon lifetime vs. temperature plots. The latter is calculated from the phonon linewidth and is an expression of Heisenberg uncertainty principle \cite{Sun2013}. The spectroscopic signatures of {\textcolor{black}{$T_{\rm g}$}} tend to appear  below 200 cm$^{-1}$. These modes are mostly due to out-of-plane motions of indium and copper [Table S2 and S3, Supporting information]. They are labeled with symmetries of the $Cc$ space group.

Interestingly, Raman scattering spectroscopy reveals that the structural aspects of \textcolor{black}{$T_{\rm g}$}  may take place in two distinct steps. This is particularly noticeable in the contour plot  [Fig. \ref{Ts_infrared_raman}(d)] as well as Fig. S3 in the Supporting Information. 
Significant broadening of all three phonons is observed around 175 and 150 K - in addition to underlying scattering intensity consistent with order-disorder processes \cite{Vysochanskii1998}. \textcolor{black}{The increase in the background scattering across $T_{\rm g}$  may be due to critical scattering (opalescence) from the microdomains.}

Taken together, we see that while many infrared- and Raman-active phonons  drive the ferroelectric transition,  the \textcolor{black}{relaxational or glassy} transition in CuInP$_2$S$_6$ is discernible only through subtle peak shifting and mode activation. This is because \textcolor{black}{$T_{\rm g}$} is not associated with a change in the space group. Instead, our spectroscopic work  reveals local lattice distortions associated with Cu site disorder.  
These effects appear only in phonons below 200 cm$^{-1}$, suggesting that  
 \textcolor{black}{$T_{\rm g}$}  involves 
local rearrangements of the Cu$^{+}$ and In$^{3+}$  coordination environments 
(which may  impact the polarization) as well as order-disorder processes that involve freezing of the Cu$^+$ ions.
 Interestingly, the two-step character of \textcolor{black}{$T_{\rm g}$} [Fig. \ref{Ts_infrared_raman}(d) and Fig. S3 in the Supporting Information] is consistent with the low frequency dielectric dispersion of CuInP$_2$S$_6$  which shows a broad transition region near 155 K along with evidence for a possible two-step aspect to the relaxation embedded in the overall shape of the response 
 at certain frequencies~\cite{Dziaugys2010}. In the $M$PS$_3$ family of materials ($M$ = Mn, Fe, Ni), only MnPS$_3$ hosts a comparable process with spin-phonon coupling across the antiferromagnetic ordering transition \cite{Vaclavkova2020}.

\section*{Summary and Outlook}

In order to explore the properties of complex chalcogenides, we measured the infrared absorption and Raman scattering response of CuInP$_2$S$_6$ across the ferroelectric and {\textcolor{black}{glassy}} phase transitions and compared our findings with a symmetry analysis and complementary lattice dynamics calculations. Several  different infrared- and Raman-active modes drive 
the ferroelectric transition whereas the {\textcolor{black}{relaxational or glassy}} phase transition involves local lattice distortions and is characterized by much more subtle peak shifting and {\textcolor{black}{activation}}. This is because $T_{\rm C}$ is a displacive transition whereas {\textcolor{black}{$T_{\rm g}$}} is due to
Cu site disorder.
Both transitions have large hysteresis regions - probably on account of the bimetallic nature of this system. The {\textcolor{black}{glassy}} transition is especially interesting because, while there is no change in space group, order-disorder processes lead to local structure differences at the Cu and In sites that arise from varying coordination environments. In addition to providing evidence for a two-step transition, the spectral response across \textcolor{black}{$T_{\rm g}$} is hysteretic and has significant underlying scattering intensity - consistent with glassy character.  
We also investigated the tendency toward chemical phase separation in these materials  and, using a combination of optical microscopy, piezoforce microscopy, transmission electron microscopy, and vibrational spectroscopies, we unravel the signature of the highest quality crystals. 
This work provides guidance on how to identify high quality single crystals and, at the same time, places the vibrational properties of ferroelectric CuInP$_2$S$_6$ on a firm foundation to support future work to reveal the symmetry and dynamics of few- and single-layer systems. 

\section*{Acknowledgements}
Research at the University of Tennessee is supported by the U.S.
Department of Energy, Office of Basic Energy Sciences, Materials
Science Division under award DE-FG02-01ER45885.  Work at Rutgers University is funded by the NSF-DMREF program (DMR-1629059) and ONR grants N00014-16-1-2951 and N00014-19-1-2073. Work at Pohang University of Science and Technology was supported by the National Research Foundation of Korea (NRF) funded by the Ministry of Science and ICT (No. 2016K1A4A4A01922028). 
This work was supported by the U.S. Department of Energy (DOE), Office of Science, Basic Energy Sciences under award
DE-SC0020353 (S. Singh).

\bibliographystyle{PRB.bst}

\end{document}